\documentclass[conference,letterpaper,10pt]{IEEEtran}
\usepackage{amsmath}
\usepackage{booktabs}
\usepackage{array}
\usepackage{graphicx}
\usepackage{multirow}
\usepackage{cite}
\usepackage{xcolor}
\usepackage{fancyhdr}
\usepackage{algorithm}
\usepackage{stfloats}
\usepackage{placeins}
\usepackage{needspace}
\usepackage{algpseudocode}

\usepackage[colorlinks=true,citecolor=black,linkcolor=black,urlcolor=black]{hyperref}
\newcommand{\panelref}[2]{\hyperref[#1]{\ref*{#1}(#2)}}
\title{Preserving Admission Responsibility in Multi-Tenant Large Language Model Prefix Caches}
\author{
  \IEEEauthorblockN{Zhiyu Wang and Rajkumar Buyya}
  \IEEEauthorblockA{\textit{Quantum Cloud Computing and Distributed Systems (qCLOUDS) Lab} \\
  \textit{School of Computing and Information Systems} \\
  \textit{The University of Melbourne, Australia} \\
  \{zhiyu.wang4, rbuyya\}@unimelb.edu.au}
}
\begin{document}
\maketitle

\begin{abstract}
Shared prefix caching turns Graphics Processing Unit (GPU) memory into persistent state shared across Large Language Model (LLM) tenants. A group that materializes new Key-Value (KV) blocks can force another to lose reusable state, yet request-time schedulers account for transient service, replacement policies primarily rank object value, and static partitioning strands idle capacity. We call this mismatch the \emph{admission-responsibility gap}. To close it, we propose PrefixShield, which meters newly materialized full KV blocks, carries responsibility across requests, gates reuse promotion while debt remains, and uses projected debt to select the group supplying eviction candidates. We implement PrefixShield in vLLM. In paired runs under one-touch pollution, PrefixShield improves victim cache hit ratio by 9.39 percentage points over the Least Recently Used (LRU) policy and 8.64 points over S3-FIFO, restoring the victim from 4.92\% to 84.87\% at 4096-block scale, and gains 2.00 points over S3-FIFO under two-pass replay. It preserves benign ShareGPT behavior and work-conserving access to idle capacity. Delayed replay yields a 35.16-point advantage while debt remains. These results show that object-value signals rank what to retain, while persistent responsibility determines which group bears reclamation pressure.
\end{abstract}

\begin{IEEEkeywords}
Large language model serving, Prefix caching, Cache replacement, Multi-tenant resource management
\end{IEEEkeywords}

\section{Introduction}
\label{sec:introduction}

Networked multi-tenant Large Language Model (LLM) services increasingly share accelerator-backed state across independent clients. Prefix caching turns Graphics Processing Unit (GPU) memory into a persistent shared resource: when requests reuse system prompts, templates, conversation history, retrieved context, or other common prefixes, retaining their Key-Value (KV) states avoids repeated prefill computation and can reduce serving cost and latency. Systems such as vLLM and SGLang retain reusable KV state across requests, while newer systems span stateful conversations, distributed context caches, multi-GPU KV placement, and multi-tier KV hierarchies~\cite{kwon2023pagedattention,zheng2024sglang,yu2025pensieve,gao2025onlinecontext,liu2025mell,qin2025mooncake}. Production-trace characterization further shows that KV-cache reuse and eviction behavior are strongly workload dependent~\cite{wang2025kvcachewild}. Global sharing is attractive because an active accounting group can exploit capacity that would otherwise sit idle.

Persistence changes the isolation problem. An accounting group can continuously submit previously unseen long prefixes, causing new KV blocks to enter the shared cache. Once capacity is exhausted, these admissions force cached state to be reclaimed, yet the group that creates the pressure is not necessarily the group whose state is evicted. A workload that repeatedly uses a stable hot prefix can therefore lose useful cached state because another group continuously introduces one-shot prefixes. Unlike bandwidth or request-time compute, cached KV state remains after the creating request completes and continues to shape later replacement decisions. This persistent-state coupling is distinct from request-time service-fairness mechanisms that account for processed client work while requests are being served~\cite{sheng2024fairness}.

We call this mismatch the \emph{admission-responsibility gap}: control over shared-state creation is decoupled from responsibility for the replacement cost that creation later induces. Existing controls leave a missing operating point. A global Least Recently Used (LRU) policy preserves work conservation, but once state is admitted, replacement ignores which accounting group created the pressure~\cite{wang2025kvcachewild}. Reuse-aware replacement policies such as S3-FIFO use quick demotion and reuse signals to estimate object value, but still decide primarily which cached objects are worth retaining~\cite{yang2023s3fifo}. Static per-group quotas provide a simple hard-isolation endpoint but can strand idle capacity, reflecting the cache-isolation and resource-limiting design space studied in multi-tenant caching~\cite{pu2016fairride,wu2022nyxcache}. Cache-admission policies decide whether an object appears valuable enough to enter the cache~\cite{einziger2017tinylfu}; they do not determine who remains accountable for pressure after admission. What is missing is a globally shared cache that remains freely borrowable while preserving the causal relationship between state creation and later reclamation.

PrefixShield targets this operating point with a simple principle: the accounting group that creates new persistent cache pressure should remain responsible for that pressure until the corresponding admission debt recovers. PrefixShield realizes this principle in three steps. First, it meters actual state creation rather than request activity, charging only newly materialized full KV blocks. Second, it prevents replay from laundering outstanding responsibility into higher retention by gating reuse promotion while debt remains. Third, when reclamation is necessary, it separates \emph{who should bear reclamation} from \emph{which of that group's blocks should be removed}: responsibility selects the group, while reuse segmentation protects more valuable state within that group. PrefixShield never denies cache admission merely because a group is in debt and imposes no fixed per-group occupancy ceiling; responsibility matters only when shared capacity becomes contested.

We implement PrefixShield in vLLM's prefix-cache management path~\cite{kwon2023pagedattention} without modifying model computation or KV tensor contents. Comparisons with S3-FIFO, a strong reuse-aware replacement policy, isolate the additional protection provided by persistent group responsibility under direct one-touch churn and two-pass replay, where object-local reuse evidence does not identify the group repeatedly creating reclamation pressure. PrefixShield also preserves work conservation: when a peer is idle, an active group occupies 509--510 of the 511 usable cache blocks. Targeted experiments span an eightfold replacement-domain range, sweep measured new-block creation relative to configured refill, and delay replay across debt clearance, testing robustness across capacity scales and isolating when responsibility becomes consequential.

This work makes three contributions:
\begin{itemize}
\item We identify the \emph{admission-responsibility gap} in multi-tenant LLM prefix caching: persistent cache pressure can be created by one accounting group while its replacement cost is externalized to others.
\item We design and implement PrefixShield, which preserves state-creation responsibility across admission, reuse, and reclamation without static cache partitioning or admission denial.
\item We show that persistent group responsibility provides protection beyond reuse-aware replacement under direct churn and replay. Against S3-FIFO, a strong reuse-aware replacement policy, PrefixShield improves victim cache-hit ratio by 8.64 percentage points under direct one-touch churn and by 2.00 percentage points under two-pass replay, while preserving benign cache-hit behavior and work-conserving access to idle capacity. Under one-touch churn at 4096-block scale, PrefixShield restores the victim hit ratio from 4.92\% under LRU to 84.87\%.
\end{itemize}

The rest of the paper is organized as follows. Section~\ref{sec:background} formulates the admission-responsibility gap and states the threat model. Section~\ref{sec:design} presents the PrefixShield design and Section~\ref{sec:implementation} its vLLM implementation. Section~\ref{sec:evaluation} evaluates protection, benign behavior, scaling, and overhead. Section~\ref{sec:discussion} discusses implications, Section~\ref{sec:related} reviews related work, and Section~\ref{sec:conclusion} concludes the paper and outlines future work.

\section{Background and Problem}
\label{sec:background}

\subsection{Shared Prefix Caching}
Autoregressive LLM serving repeatedly performs the same prefix computation when requests share system prompts, templates, conversation history, retrieved context, or other common token sequences. Prefix caching retains the KV states produced during prefill and reuses them when a later request matches an existing prefix, as in block-based, radix-tree, and stateful-conversation serving designs~\cite{kwon2023pagedattention,zheng2024sglang,yu2025pensieve}. Recent systems additionally fuse cached knowledge, distribute context caches, migrate KV state across GPUs, or hierarchically place KV state to reduce memory and data-movement costs~\cite{yao2025cacheblend,gao2025onlinecontext,liu2025mell,qin2025mooncake}. Because GPU memory is finite, cached blocks eventually compete for capacity and must be reclaimed before new KV state can be materialized; production traces confirm that reuse patterns and eviction-policy effectiveness vary substantially across workloads~\cite{wang2025kvcachewild}. A global replacement domain is naturally work-conserving, but it also couples accounting groups through reclamation: state created by one group can determine which other group's reusable state is displaced.

\subsection{The Admission-Responsibility Gap}
The central problem is a mismatch between who creates cache pressure and who bears its consequences. Consider a victim that repeatedly accesses a stable hot prefix while another accounting group continuously submits previously unseen long prefixes. Each such request materializes new KV blocks. Under global LRU, those blocks enter the same replacement domain as the victim's useful state, so sustained admissions can displace the victim even when the newly introduced prefixes are never reused. The group generating new state controls admission pressure, but the associated eviction cost is externalized to the shared cache. This cross-client coupling is related to long-standing multi-tenant cache-sharing concerns, but here the shared objects are expensive-to-reconstruct KV states created by inference requests~\cite{pu2016fairride,wang2025kvcachewild,li2025oneiros}.

This distinction separates request activity from cache creation. A request that hits a long cached prefix may generate almost no new KV state, whereas a similarly sized request containing an unseen suffix may materialize many new blocks. Likewise, a group need not generate a conspicuous burst: continuously rotating unseen prefixes at a lower rate can accumulate persistent pressure over time. A mechanism that targets this gap must therefore account for actual state creation rather than infer pressure solely from request rate.

Two design endpoints illustrate the missing operating point. Global replacement provides work conservation without accounting-group-aware protection; hard quotas provide structural isolation but strand capacity when some groups are idle. PrefixShield targets the point between them: a globally shared cache in which responsibility for creating pressure remains relevant when reclamation is required.

\subsection{Threat Model and Design Goals}
We consider a multi-tenant LLM serving system in which requests from multiple operator-defined accounting groups share a finite physical KV-block pool and one replacement domain. The adversary is a legitimate client that can control the token sequences and timing of its own requests. It may generate unseen prefixes, rotate them over time, or replay previously admitted prefixes. It cannot directly modify another group's requests, cache metadata, or KV contents and does not compromise the model, runtime, operating system, or authentication layer. Network flooding, compute-only denial of service, memory-safety exploits, and unbounded creation of independently entitled accounting groups are outside the scope of this work.

In the prototype and evaluation, native prefix lookup is namespace-scoped by accounting group even though all groups compete for the same physical block pool. Cross-group interference therefore arises through shared capacity and replacement rather than cross-group content deduplication. A managed block is assigned to the accounting group that materializes it, and this ownership remains fixed for the block's lifetime; requester identity is tracked separately. Cross-group content reuse would require multi-principal rules for ownership and admission responsibility and is outside the current prototype.

We exercise three complementary attack patterns grounded in prior cache and LLM-serving studies. \emph{One-touch pollution}~\cite{wang2025kvcachewild} continuously introduces unseen prefixes. \emph{Low-and-slow rotation}~\cite{xiang2026servegen} spreads similar pressure over time. \emph{Replay-based pollution}~\cite{yu2025iccache,einziger2017tinylfu} revisits previously admitted prefixes in an attempt to manufacture apparent reuse. These workloads stress direct state creation, temporal evasion, and reuse laundering, respectively.

PrefixShield has four goals: (G1) attribute sustained cache pressure so that one group cannot freely externalize its replacement cost; (G2) preserve responsibility across time and prevent unresolved pressure from being erased through replay; (G3) retain work conservation so idle capacity remains globally borrowable; and (G4) limit added overhead on the cache-management path. 

\begin{figure*}[!t]
\centering
\includegraphics[width=0.98\textwidth]{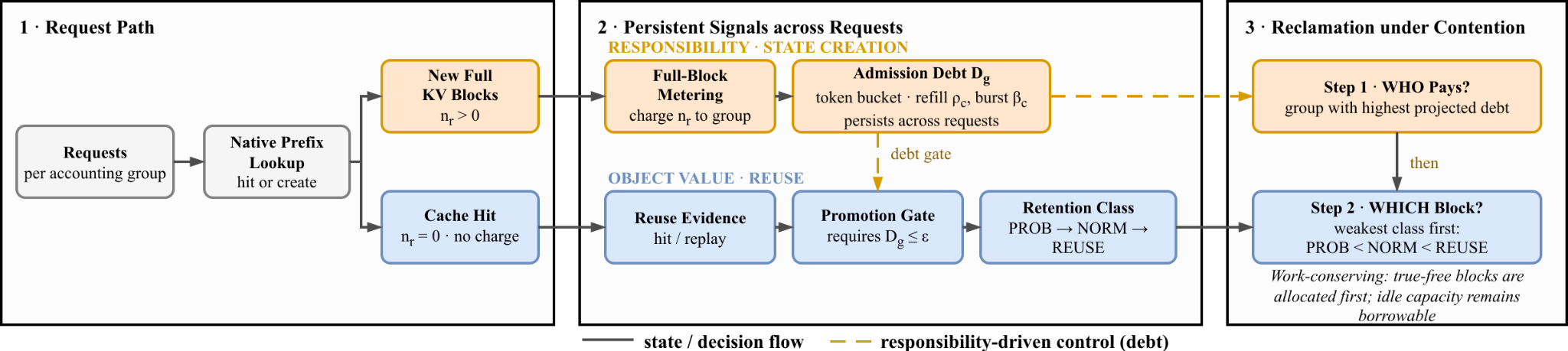}
\caption{PrefixShield's request-to-reclamation lifecycle. New KV blocks accrue admission responsibility; cache hits provide reuse evidence. Debt gates promotion across requests and, under contention, selects who pays, while retention class selects which block is reclaimed.}
\label{fig:prefixshield-overview}
\end{figure*}

\section{PrefixShield}
\label{sec:design}

PrefixShield is based on a simple principle: the accounting group that introduces new shared-cache pressure should retain responsibility for that pressure until its admission debt recovers. Responsibility does not prevent use of spare capacity; it becomes consequential only when capacity is contested.

\subsection{Design Overview}
Figure~\ref{fig:prefixshield-overview} summarizes PrefixShield's three-stage request-to-reclamation lifecycle and the two persistent signals that connect them. On the request path, native prefix lookup separates reuse of already resident state from creation of new persistent state. Cached blocks incur no new admission charge, whereas every newly materialized full KV-cache block is charged to the accounting group that caused its creation. Across requests, admission debt preserves creation responsibility while reuse evidence captures object value: outstanding debt gates reuse promotion, and when shared capacity becomes contested, projected debt selects the responsible group before retention class selects the victim block.

Each managed block has an immutable accounting owner and retention class $\kappa(b)\in\{\textsc{Probationary},\textsc{Normal},\textsc{Reuse}\}$. The logical prefix identifier used for admission and reuse history is policy metadata, distinct from vLLM's native content-hash lookup key; namespace isolation is carried separately through the native cache namespace. Newly admitted state is \textsc{Probationary} on its first successful admission or whenever its admission leaves the responsible group in debt; a funded subsequent admission of an already admitted prefix may enter \textsc{Normal}. Proven reuse can promote matching state to \textsc{Reuse}, but only after the responsible accounting group's effective debt has cleared. Touches and promotion update access state but do not migrate ownership. Reclamation is two-stage: debt determines \emph{which group bears reclamation}, then the retention class determines \emph{which block owned by that group is reclaimed}.

Two invariants capture the design. \textbf{I1: Complete pressure attribution.} Every newly materialized full KV-cache block is charged exactly once to the accounting group that introduced it, whereas pure cache reuse creates no additional admission responsibility. \textbf{I2: No debt laundering through reuse.} Reaccessing a prefix may establish reuse evidence, but outstanding admission debt prevents that evidence from immediately promoting state to the \textsc{Reuse} class.

\textbf{Design rationale.} PrefixShield keeps creation responsibility, reuse value, and physical contention as separate signals. Request rate is not a reliable proxy for new state creation, reuse alone can be manufactured through replay, and hard enforcement of responsibility would sacrifice borrowing when capacity is idle. PrefixShield therefore meters creation at materialization, interprets reuse through the debt gate, and consults responsibility only when reclamation is actually required.

\subsection{Fully Metered Admission Responsibility}
PrefixShield meters cache creation rather than request arrival. Let $n_r$ be the number of full KV-cache blocks newly materialized by request $r$ after native prefix lookup. PrefixShield charges exactly $n_r$ blocks to the request's accounting group and service contract; therefore a pure cache hit has $n_r=0$ and creates no new charge. This distinction is essential because two requests of similar token length can create very different amounts of persistent shared state.

For each accounting group $g$ and service contract $c$, PrefixShield maintains a block-denominated balance $B_{g,c}$, positive refill rate $\rho_c$ in blocks per unit accounting time, and burst capacity $\beta_c$ in blocks. We use a monotonic accounting clock $t$; lazy refill requires no background timer. If the stored balance was last updated at time $t_0$, its effective value at time $t$ is
\begin{equation}
\widehat{B}_{g,c}(t)=\min\!\left\{\beta_c,\,B_{g,c}(t_0)+\rho_c(t-t_0)\right\}.
\label{eq:effective-balance}
\end{equation}
When request $r$ creates $n_r$ new full blocks, PrefixShield stores
\begin{equation}
B_{g,c}(t)=\widehat{B}_{g,c}(t)-n_r.
\label{eq:meter-charge}
\end{equation}
A negative balance does not suppress cache admission. It represents outstanding admission responsibility while preserving work conservation. The effective raw debt of accounting group $g$ at time $t$ is
\begin{equation}
D_g(t)=\sum_c \max\!\left\{0,-\widehat{B}_{g,c}(t)\right\}.
\label{eq:raw-debt}
\end{equation}
Throughout, $\epsilon\ge0$ is the implementation tolerance for treating debt as cleared. Because $\widehat{B}_{g,c}(t)$ is evaluated lazily, debt recovers with accounting time when a group stops creating new state; eviction itself does not repay the token bucket. The parameters $\beta_c$ and $\rho_c$ express contract-level responsibility rather than admission control: $\rho_c$ is the service tier's sustainable new-block creation entitlement, and $\beta_c$ is its short-term burst credit. Operators can derive $\rho_c$ from tier-level state-creation budgets and choose $\beta_c$ for a desired burst horizon. Neither parameter imposes a cache-occupancy quota or an upper bound on request-driven state creation.

\textbf{Accounting principal versus service contract.} The accounting group is the administrative principal to which persistent responsibility attaches, while the service contract specifies that principal's creation entitlement. Raw debt is aggregated at the group level for reuse gating so contract relabeling cannot erase unresolved responsibility; contract-specific refill rates and normalized weights instead determine how heterogeneous entitlements influence reclamation priority.

\noindent\textbf{Proposition 1 (Quiescent debt clearance).} Assume $\rho_c>0$ for every active service contract. Fix group $g$ at logical time $t_0$, let $d_{g,c}(t_0)=\max\{0,-\widehat{B}_{g,c}(t_0)\}$, and define $\tau_g(t_0)=\max_{c:d_{g,c}(t_0)>0}d_{g,c}(t_0)/\rho_c$, taking the maximum of the empty set as zero. If $g$ materializes no new blocks after $t_0$, then $D_g(t)=0$ for all $t\ge t_0+\tau_g(t_0)$. \emph{Proof.} Under quiescence, contract $c$'s deficit is $\max\{0,d_{g,c}(t_0)-\rho_c(t-t_0)\}$ until it reaches zero. Every contract deficit therefore vanishes by $t_0+\tau_g(t_0)$, and Eq.~\eqref{eq:raw-debt} proves the claim.

The post-charge balance and prefix-admission history determine the initial class. A first successful admission is \textsc{Probationary}. If $B_{g,c}(t)<-\epsilon$, newly admitted blocks are also \textsc{Probationary} regardless of prior history. Otherwise, a funded subsequent admission of a prefix already admitted by the same accounting group may enter \textsc{Normal}. PrefixShield therefore changes retention exposure rather than denying new state.

\subsection{Debt-Gated Reuse Promotion}
Reuse is valuable evidence, but reuse alone is an unsafe protection signal. After introducing substantial state, an adaptive group could replay those prefixes and attempt to convert self-generated pollution into higher-retention cache entries. PrefixShield therefore separates observing reuse from granting reuse privilege. A cached block is promotion-eligible only when the current request represents proven reuse and the block metadata matches the same accounting group, contract, and logical prefix. Even then, promotion to \textsc{Reuse} is granted only when the responsible group's effective raw debt has cleared:
\begin{equation}
\kappa(b)\rightarrow\textsc{Reuse}
\Longrightarrow
\mathrm{ProvenReuse}(b,r)\ \land\ D_g(t)\le\epsilon.
\label{eq:promotion-gate}
\end{equation}
When $D_g(t)>\epsilon$, PrefixShield still records reuse and updates access metadata, but the block remains in its current class. Because $D_g$ aggregates debt over all contracts of the same accounting group, switching contract labels within that group cannot bypass unresolved admission responsibility; contract weights affect reclamation ranking, not whether outstanding responsibility may be laundered into \textsc{Reuse} promotion. Proposition~1 shows that the gate is not permanent: under quiescence, eligibility returns after outstanding deficits refill.

Algorithm~\ref{alg:admission-reuse} summarizes these request-side updates. It meters newly materialized full KV blocks, initializes their retention class, records reuse evidence, and applies the debt-gated promotion rule.

\begin{algorithm}[!t]
\caption{Admission accounting and debt-gated reuse hooks}
\label{alg:admission-reuse}
\footnotesize
\begin{algorithmic}[1]
\Procedure{OnRequest}{$r,g,p$}
    \State $q_r\gets\Call{PriorReferenceCount}{g,p}$
    \State \Call{RecordRequestReference}{$r,g,p$}
\EndProcedure
\Statex
\Procedure{OnCacheFullBlocks}{$r,g,c,p,t,n_{\mathrm{cached}},n_{\mathrm{full}}$}
    \State $n_r\gets\max(0,n_{\mathrm{full}}-n_{\mathrm{cached}})$
    \If{$n_r>0$}
        \State $\widehat{B}_{g,c}(t)\gets\Call{RefillEffectiveBalance}{g,c,t}$
        \State $B_{g,c}(t)\gets\widehat{B}_{g,c}(t)-n_r$
        \If{$B_{g,c}(t)<-\epsilon$ \textbf{or} $(g,p)$ has no prior successful admission}
            \State $\kappa_{\mathrm{new}}\gets\textsc{Probationary}$
        \Else
            \State $\kappa_{\mathrm{new}}\gets\textsc{Normal}$
        \EndIf
    \EndIf
    \State \Call{NativeCacheFullBlocks}{$r$}
    \If{$n_r>0$}
        \State \Call{RecordSuccessfulAdmission}{$g,p$}
        \State \Call{AttachBlockMetadata}{$r,g,c,p,\kappa_{\mathrm{new}}$}
    \EndIf
\EndProcedure
\Statex
\Procedure{OnTouch}{$r,g,c,p,t,H$}
    \State $\mathrm{proven}\gets(q_r\ge1)\land((g,p)\text{ was admitted})$
    \State $d\gets D_g(t)$
    \ForAll{$b\in H$ with PrefixShield metadata}
        \State \Call{UpdateAccessMetadata}{$b$}
        \If{$\mathrm{proven}$ \textbf{and} $b$ matches $(g,c,p)$ \textbf{and} $\kappa(b)\neq\textsc{Reuse}$}
            \If{$d\le\epsilon$} \State $\kappa(b)\gets\textsc{Reuse}$
            \Else \State \Call{RecordDebtGatedReuse}{$b$}
            \EndIf
        \EndIf
    \EndFor
\EndProcedure
\end{algorithmic}
\end{algorithm}

\subsection{Responsibility-Aware Reuse-Segmented Eviction}
When shared capacity becomes contested, PrefixShield separates two decisions that global replacement collapses into one: \emph{which accounting group should bear reclamation} and \emph{which block owned by that group should be evicted}. Admission debt answers the first question; reuse segmentation answers the second.

Let $\rho_{\min}=\min_{c'}\rho_{c'}$ and $w_c=\rho_c/\rho_{\min}$. The normalized debt of accounting group $g$ is
\begin{equation}
\widetilde{D}_g(t)=\sum_c\frac{\max\!\left\{0,-\widehat{B}_{g,c}(t)\right\}}{w_c}.
\label{eq:normalized-debt}
\end{equation}
This normalization measures excess pressure relative to contract entitlement. For equal raw deficits, a contract with larger $\rho_c$ contributes less normalized debt because its service tier permits a higher sustained creation rate; the debt remains fully metered, while reclamation priority reflects pressure relative to the configured contract. For candidate set $C$, the indebted candidate-owning groups are
\begin{equation}
\mathcal{G}^{+}(C,t)=\left\{g:\exists b\in C,\ \mathrm{owner}(b)=g,\ \widetilde{D}_g(t)>\epsilon\right\}.
\label{eq:eligible-groups}
\end{equation}
During a multi-block allocation, the selector maintains a temporary projected normalized responsibility score $P_g^{(j)}$, initialized as $P_g^{(0)}=\widetilde{D}_g(t)$. Let $C_j$ be the candidates remaining at step $j$, let $W_g(t)$ be the maximum weight among group $g$'s currently indebted contracts, and define $\mathcal{G}^{+}_j=\{g:\exists b\in C_j,\ \mathrm{owner}(b)=g,\ P_g^{(j)}>\epsilon\}$. At victim-selection step $j$, responsibility is assigned to
\begin{equation}
g_j^{*}=\arg\max_{g\in\mathcal{G}^{+}_j}P_g^{(j)},
\label{eq:responsible-group}
\end{equation}
with equal projected debt broken by retention segment and native queue order. After one victim is selected from $g_j^{*}$,
\begin{equation}
P_{g_j^{*}}^{(j+1)}=\max\!\left\{0,P_{g_j^{*}}^{(j)}-\frac{1}{W_{g_j^{*}}(t)}\right\}.
\label{eq:projected-normalized}
\end{equation}
Since $1/W_g(t)$ is the smallest normalized one-block decrement among group $g$'s currently indebted contracts, this contract-agnostic update conservatively reduces projected responsibility without assuming which contract a reclaimed block should repay. The projection affects only victim selection within the current allocation batch; persistent token-bucket balances evolve solely through new-state metering and refill.

Within the responsible group, candidate blocks follow
\begin{equation}
\textsc{Probationary}\prec\textsc{Normal}\prec\textsc{Reuse},
\label{eq:segment-order}
\end{equation}
with native free-queue order as the recency tie-break inside each segment. If no candidate-owning group has positive projected debt, selection falls back to the same segment ordering globally. Thus, debt determines \emph{who pays}, while reuse segmentation determines \emph{which block pays}.

Algorithm~\ref{alg:reclamation} summarizes reclamation under contention. It allocates truly free blocks first; if additional space is required, projected admission debt selects the responsible group, and retention class selects the victim block within that group.

\begin{algorithm}[!t]
\caption{Responsibility-aware batch reclamation}
\label{alg:reclamation}
\footnotesize
\begin{algorithmic}[1]
\Require Free-queue candidates $C$, requested blocks $k$, logical time $t$
\State $S\gets\Call{TakeTrueFree}{C,k}$; $m\gets k-|S|$
\If{$m=0$} \State \Return $S$ \EndIf
\State $Q\gets\Call{SegmentQueuesByOwner}{C\setminus S}$
\State $(\widetilde{D},W)\gets\Call{EffectiveDebtContext}{t}$; $P\gets\widetilde{D}$
\While{$m>0$}
    \State $\mathcal{G}^{+}\gets\{g:Q[g,\cdot]\neq\emptyset\land P_g>\epsilon\}$
    \If{$\mathcal{G}^{+}\neq\emptyset$}
        \State $g^{*}\gets\Call{ArgMaxDebt}{\mathcal{G}^{+},P,Q}$
        \State $b\gets\Call{FirstSegmentQueue}{Q[g^{*},\cdot]}$
        \State $P_{g^{*}}\gets\max\{0,P_{g^{*}}-1/W_{g^{*}}(t)\}$
    \Else
        \State $b\gets\Call{GlobalFirstSegmentQueue}{Q}$
    \EndIf
    \State Append $b$ to $S$; remove $b$ from $Q$; $m\gets m-1$
\EndWhile
\State \Return $S$
\end{algorithmic}
\end{algorithm}

\subsection{Work-Conserving Sharing}
PrefixShield deliberately avoids hard partitions. Admission debt does not define a maximum occupancy, and no block is rejected merely because its group has exhausted a pressure budget. When unused capacity exists, any accounting group may consume it; truly free blocks are allocated before cached state is reclaimed. Responsibility-aware replacement activates only when shared capacity becomes contested. Together, these mechanisms preserve native cache admission while modifying only retention classification, promotion eligibility, and victim choice.

\section{Implementation}
\label{sec:implementation}
We implement PrefixShield on top of vLLM's V1 prefix-caching path~\cite{kwon2023pagedattention}. The prototype preserves native prefix lookup, KV-block allocation, cache insertion, and block-release semantics and augments the cache-management path with responsibility accounting and replacement metadata. PrefixShield does not modify model weights, attention kernels, KV tensor contents, or request-visible generation semantics.

\textbf{Integration points.} PrefixShield observes request allocation, block materialization after prefix matching, cache-touch events, and cached-block reclamation. The allocation path associates each request with its accounting group, service contract, logical prefix identifier, and accounting time. The evaluation maps each accounting group to a native cache namespace while all groups continue to compete for the same physical replacement domain. Materialization charges only full blocks that are actually new; touches update access metadata and execute debt-gated promotion; reclamation replaces the default victim choice with responsibility-aware segmented selection while retaining native queue order inside each class.

\textbf{Metadata and selector.} Each managed block records an immutable accounting owner together with its service contract, logical prefix identifier, retention class, prefix depth, insertion order, recent access order, and access count. Per group/contract accounting state consists of a block-denominated balance and last refill time. Prefix-reference and successful-admission history distinguish first admission from established reuse, while requester fields remain separate from owner fields during reclamation and touch processing. Metadata is maintained outside the KV payload and does not alter native KV contents or block hashes. The prototype selector scans reclaimable candidates in software for auditability. A production implementation can maintain indexed per-group and per-segment queues to reduce selection work without changing policy semantics; Section~\ref{sec:evaluation} therefore measures selector latency and metadata footprint rather than assuming negligible overhead.

\section{Performance Evaluation}
\label{sec:evaluation}
We evaluate whether PrefixShield (i) protects victim KV-cache state under direct and adaptive pollution, (ii) preserves benign performance and work-conserving sharing, and (iii) remains effective across cache scale, creation pressure, nominal load, and deployment settings. We then measure its overhead and isolate the contributions of reuse segmentation and debt gating.

\subsection{Experimental Setup}
\textbf{System and platforms.} We implement PrefixShield in vLLM~0.10.2 with Automatic Prefix Caching~\cite{kwon2023pagedattention}. The default platform uses Qwen2.5-1.5B-Instruct on an NVIDIA L4. Portability runs use NVIDIA A100 with Qwen2.5-1.5B-Instruct and Qwen2.5-7B-Instruct.

\textbf{Cache and policy configuration.} Unless otherwise stated, the default configuration uses 512 configured GPU KV-cache blocks, of which 511 are usable because vLLM reserves one block. It uses 16-token blocks, 1,536-token shared prefixes, and the 60\% nominal load profile. The replacement domain is deliberately capped to create controlled, repeatable contention within finite traces; the configured block count is an experimental cache-capacity parameter rather than an estimate of the GPU's maximum KV capacity. PrefixShield uses refill rates of 64, 40, and 16 blocks/s and burst capacities of 256, 192, and 128 blocks for critical, standard, and best-effort contracts, respectively. This policy configuration is fixed across the reported workloads; the sensitivity experiments vary workload pressure and deployment conditions rather than retuning PrefixShield per trace. For each deployment profile, $C_{\mathrm{prefill}}$ is the LRU unique-prefill throughput calibrated from 32 distinct-prefix requests. A nominal load factor $\alpha$ sets the base arrival rate to $\alpha C_{\mathrm{prefill}}$; the one-touch pressure phase uses $\min\{0.75C_{\mathrm{prefill}},1.15\alpha C_{\mathrm{prefill}}\}$.

\textbf{Baselines.} We compare PrefixShield against three complementary cache-management baselines representing distinct design points:
\begin{itemize}
\item \textbf{LRU}~\cite{wang2025kvcachewild}: vLLM's global work-conserving replacement, allowing unrestricted borrowing of idle cache capacity.
\item \textbf{S3-FIFO}~\cite{yang2023s3fifo}: a work-conserving quick-demotion policy with equal-size KV objects, a 10\% small First-In, First-Out (FIFO) queue, a 90\% main FIFO queue, an equal-size ghost FIFO queue, and a capped two-bit counter; it receives no tenant or responsibility metadata.
\item \textbf{StaticQuota}~\cite{pu2016fairride,wu2022nyxcache}: a fixed equal per-group partition baseline representing hard isolation; it permits no cross-group borrowing.
\end{itemize}
Together, they span native global sharing, reuse-aware object-value replacement, and hard isolation. Main attack, benign, and borrowing experiments report all four policies; ablations disable individual PrefixShield components.

\textbf{Workloads.} We evaluate six primary traces spanning three workload categories:
\begin{itemize}
\item \textbf{Direct and adaptive pollution:} The \emph{one-touch} trace~\cite{wang2025kvcachewild} continuously introduces unseen prefixes, reflecting low-reuse admission pressure observed in production KV-cache behavior. The \emph{low-and-slow} trace~\cite{xiang2026servegen} spreads similar pressure over time. The \emph{two-pass replay} trace~\cite{yu2025iccache,einziger2017tinylfu} revisits previously admitted prefixes to stress reuse-aware retention while creation responsibility remains outstanding. Each attack trace uses four 1,536-token victim hot prefixes, corresponding to 96 blocks per prefix and 384 blocks in aggregate under the default 16-token block size.
\item \textbf{Benign traffic:} The \emph{ShareGPT} trace~\cite{zheng2024sglang} represents conversational prefix reuse. The \emph{bursty ShareGPT} trace~\cite{zheng2024sglang,xiang2026servegen} concentrates benign arrivals into short bursts.
\item \textbf{Work conservation:} The controlled \emph{borrowing} trace~\cite{pu2016fairride,wu2022nyxcache} leaves one accounting group idle and measures how much of the shared cache an active peer can occupy.
\end{itemize}
The three attack traces vary how cache pressure is expressed through continuous creation, temporally dispersed creation, and replay. The two benign traces test compatibility with conversational reuse under ordinary and burst-concentrated arrivals. The borrowing trace tests whether protection preserves access to otherwise idle cache capacity.

\textbf{Methodology.} Reported paired comparisons use 10 matched runs. Within each run pair, compared policies receive identical request traces and workload parameters. For request $r$, cache-hit fraction is $h_r=\texttt{cached\_tokens}_r/\texttt{prompt\_tokens}_r$, as reported by vLLM after native prefix lookup. We report run-level mean cache-hit fractions over measured victim requests for attack traces and over all measured requests for benign traces. The 95th Percentile (P95) of Time to First Token (TTFT) is computed over the same request population. Where reported, a 95\% bootstrap Confidence Interval (CI) is computed over paired run-level differences.

\subsection{Protection under Direct and Adaptive Pollution}
\begin{table}[t]
\centering
\caption{Victim cache protection under direct and adaptive pollution.}
\label{tab:attack-results}
\small
\renewcommand{\arraystretch}{0.92}
\begin{tabular*}{\columnwidth}{@{\extracolsep{\fill}}llrr@{}}
\toprule
Workload & Policy & Hit (\%) $\uparrow$ & P95 TTFT (ms) $\downarrow$ \\
\midrule
\multirow{4}{*}{One-touch}
  & LRU & 5.54 & 1440.0 \\
  & S3-FIFO & 6.29 & 1814.8 \\
  & StaticQuota & 1.94 & 1466.1 \\
  & PrefixShield & \textbf{14.93} & \textbf{1428.2} \\
\midrule
\multirow{4}{*}{Low-and-slow}
  & LRU & 92.64 & 66.3 \\
  & S3-FIFO & 98.64 & \textbf{34.5} \\
  & StaticQuota & 0.00 & 155.3 \\
  & PrefixShield & \textbf{98.90} & 34.7 \\
\midrule
\multirow{4}{*}{Two-pass replay}
  & LRU & 92.87 & 132.9 \\
  & S3-FIFO & 93.96 & 141.3 \\
  & StaticQuota & 26.31 & 316.4 \\
  & PrefixShield & \textbf{95.97} & \textbf{104.8} \\
\bottomrule
\end{tabular*}
\end{table}

Table~\ref{tab:attack-results} summarizes victim-side protection. \textbf{One-touch pollution.} PrefixShield raises victim cache hit ratio from 5.54\% under LRU to 14.93\%, a paired gain of 9.39 Percentage Points (pp; 95\% CI: 7.94--10.96 pp). S3-FIFO remains at 6.29\%, leaving PrefixShield 8.64 pp higher (95\% CI: 7.23--9.43 pp). This is the largest absolute protection gap among the three default attack workloads and most clearly separates persistent responsibility from quick demotion under direct one-shot churn. Absolute victim hit ratios remain low at this scale because the default configuration is deliberately extreme: the 384-block victim hot set occupies 75\% of the 511 usable blocks, so even protected victims face frequent reclamation; under the same attack window at 4096 blocks, PrefixShield restores the victim to 84.87\% versus 4.92\% for LRU (Fig.~\panelref{fig:sensitivity}{a}). StaticQuota reaches 1.94\% because its 255-block victim partition is smaller than the 384-block hot set: fixed isolation does not help when the entitled share is itself smaller than the working set. Mean P95 TTFT remains comparable, at 1440.0\,ms under LRU and 1428.2\,ms under PrefixShield, indicating that the cache-protection gain does not come at the expense of tail latency.

\textbf{Low-and-slow rotation.} Without attack, the LRU control retains 99.19\% of the victim prefix; slow rotation lowers attacked LRU to 92.64\%, a 6.56-pp mean residency loss. PrefixShield restores the attacked hit ratio to 98.90\%, recovering 6.26 pp, or 95.5\% of that mean loss. S3-FIFO reaches 98.64\%, while PrefixShield retains a paired 0.25-pp advantage (95\% CI: 0.06--0.51 pp). StaticQuota falls to 0.00\% because its 255-block victim partition is smaller than the reuse distance across the four 96-block hot prefixes: the three intervening prefixes occupy 288 blocks, causing each prefix to be evicted before its next access. Quick demotion therefore removes most low-reuse pressure in this regime, while PrefixShield retains a statistically positive 0.25-pp advantage and extends responsibility-aware protection to the direct-churn and replay regimes examined next.

\textbf{Two-pass replay.} The no-attack LRU control again retains 99.19\%. Under replay, LRU, S3-FIFO, and PrefixShield reach 92.87\%, 93.96\%, and 95.97\%, respectively. PrefixShield gains 3.10 pp over attacked LRU and exceeds S3-FIFO by 2.00 pp (95\% CI: 0.93--3.23 pp), while reducing P95 TTFT from 141.3 to 104.8\,ms relative to S3-FIFO. Revisited attacker state gains additional reuse evidence, yet outstanding creation responsibility remains useful for reclamation; the delayed-replay ablation in Fig.~\panelref{fig:mechanism-ablation}{b} isolates the debt-gating mechanism behind this gap, where the advantage reaches 35.16 pp while responsibility remains outstanding.

\textbf{Takeaway.} PrefixShield retains clear gains over reuse-aware replacement under direct one-touch churn and replay, where object-local reuse evidence does not identify the group repeatedly creating reclamation pressure. Object-value estimation reasons about the observed value of cached state; PrefixShield contributes the complementary signal of persistent group responsibility.

\subsection{Benign Performance and Work-Conserving Sharing}
\begin{table}[t]
\centering
\caption{Benign performance.}
\label{tab:benign}
\small
\renewcommand{\arraystretch}{0.92}
\begin{tabular}{@{}llrr@{}}
\toprule
Workload & Policy & Cache Hit (\%) $\uparrow$ & P95 TTFT (ms) $\downarrow$ \\
\midrule
\multirow{4}{*}{ShareGPT}
 & LRU & 35.83 & 106.6 \\
 & S3-FIFO & 38.85 & \textbf{96.5} \\
 & StaticQuota & 35.24 & 110.2 \\
 & PrefixShield & \textbf{38.92} & 97.4 \\
\midrule
\multirow{4}{*}{Bursty}
 & LRU & 37.17 & 190.9 \\
 & S3-FIFO & \textbf{38.72} & 185.5 \\
 & StaticQuota & 36.46 & 205.3 \\
 & PrefixShield & 38.10 & \textbf{168.1} \\
\bottomrule
\end{tabular}
\end{table}
Table~\ref{tab:benign} summarizes benign behavior. The paired PrefixShield--S3-FIFO cache-hit differences are statistically indistinguishable on both traces: 38.92\% versus 38.85\% on ShareGPT and 38.10\% versus 38.72\% under bursty ShareGPT. Responsibility-aware protection therefore does not require a broad benign cache-hit sacrifice relative to the S3-FIFO replacement baseline; statistical indistinguishability here is the intended safety result, not evidence that PrefixShield should dominate every benign trace. On ShareGPT, PrefixShield records 97.4\,ms P95 TTFT versus 96.5\,ms for S3-FIFO; under bursty ShareGPT it records 168.1\,ms versus 185.5\,ms. In particular, the bursty workload introduces short periods of concentrated admission without causing permanent classification as harmful: responsibility decays through refill, and legitimate reuse remains eligible for normal protection after debt clears.

\begin{table}[t]
\centering
\caption{Work-conserving borrowing. Borrower Blocks denotes the observed borrower occupancy; ``--'' denotes unavailable per-group accounting.}
\label{tab:borrowing}
\small
\renewcommand{\arraystretch}{0.92}
\resizebox{\columnwidth}{!}{%
\begin{tabular}{@{}lrrr@{}}
\toprule
Policy & Cache Hit (\%) $\uparrow$ & P95 TTFT (ms) $\downarrow$ & Borrower Blocks $\uparrow$ \\
\midrule
LRU & \textbf{99.33} & \textbf{38.1} & -- \\
S3-FIFO & 92.78 & 93.6 & \textbf{509--510} \\
StaticQuota & 18.18 & 277.1 & 254--255 \\
PrefixShield & \textbf{99.33} & 38.5 & \textbf{509--510} \\
\bottomrule
\end{tabular}}
\end{table}
Table~\ref{tab:borrowing} isolates asymmetric borrowing. StaticQuota caps the active borrower at 254--255 blocks, approximately one equal share, whereas S3-FIFO and PrefixShield both occupy 509--510 of the 511 usable blocks, confirming work-conserving access to idle capacity. Native LRU is globally work-conserving but does not expose an equivalent per-group occupancy counter, so its Borrower Blocks entry is marked ``--''. PrefixShield also matches LRU's 99.33\% hit ratio. Capacity work conservation and retention efficiency are therefore distinct, and PrefixShield preserves LRU-level retention without a fixed partition. Thus, PrefixShield behaves like a globally shared cache when contention is absent while retaining responsibility-aware reclamation when contention appears.

\subsection{Scaling, Operating Boundary, Overhead, and Ablation}
\begin{figure*}[t]
\centering
\includegraphics[width=0.9\textwidth]{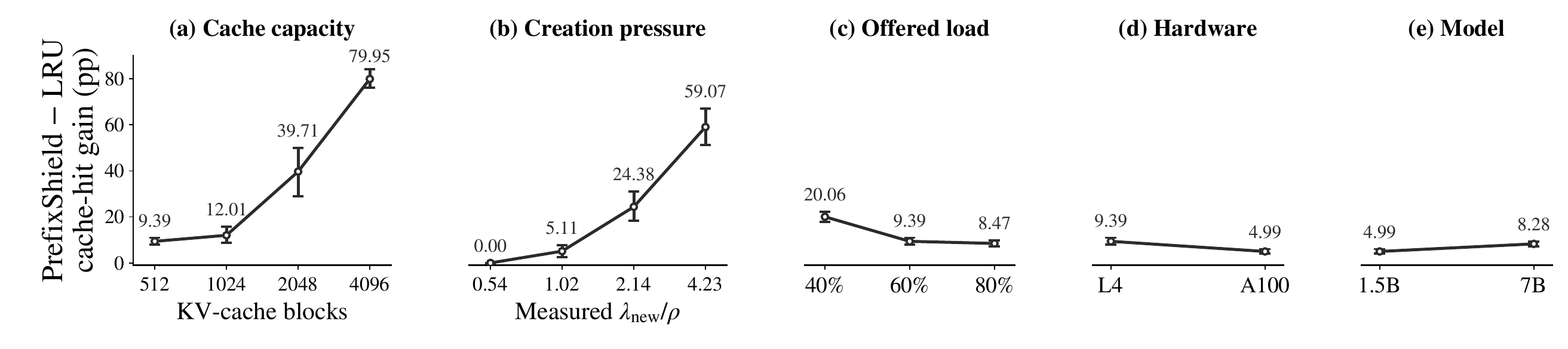}
\caption{Scaling, operating boundary, and deployment sensitivity. Panels report paired PrefixShield-minus-LRU victim cache-hit gains with 95\% bootstrap CIs: (a) cache scale, (b) measured new-block creation/refill ratio, (c) nominal load factor, (d) hardware, and (e) model deployment.}
\label{fig:sensitivity}
\end{figure*}

We use one-touch pollution for scale and deployment sensitivity because it exposes direct cache-creation pressure under a common paired protection metric. These sweeps test the effect of enabling PrefixShield as deployment conditions vary. We therefore use vLLM's unmodified LRU path as the common counterfactual. Figure~\ref{fig:sensitivity} reports paired PrefixShield-minus-LRU victim cache-hit differences across cache capacity, measured new-block creation pressure relative to refill, nominal load, hardware, and model deployment. The creation-pressure ratio is an observed workload rate divided by the configured responsibility refill rate; it is not an admission threshold because PrefixShield never rejects cache creation.

\textbf{Scale and operating boundary.} The victim working set scales with capacity, occupying approximately 76.6\% of usable blocks at 1024--4096 blocks, while the one-touch attack window remains fixed. PrefixShield's paired gain over LRU is 9.39, 12.01, 39.71, and 79.95 pp at 512, 1024, 2048, and 4096 blocks, respectively; at 4096 blocks, the absolute victim hit ratios are 4.92\% for LRU and 84.87\% for PrefixShield. Relative churn decreases with capacity, yet the 4096-block run still materializes 5.64 full-cache turnovers on average, establishing protection across an eightfold replacement-domain range. Because the debt gate and reclamation priority depend on measured creation relative to configured refill, the creation-pressure sweep equivalently characterizes sensitivity to the operator-chosen $\rho_c$ (halving every refill rate doubles the measured ratio). At a ratio of 0.54, mean debt and the PrefixShield-minus-LRU gain are both zero: creation within the configured entitlement accrues no responsibility by design. At ratios of 1.02, 2.14, and 4.23, the gains rise to 5.11, 24.38, and 59.07 pp as mean pressure debt rises to 5.69, 260.90, and 901.91 blocks. The token bucket absorbs short bursts through $\beta_c$, so a trace-averaged ratio near one does not define a discontinuous threshold; sustained creation beyond configured refill progressively increases outstanding responsibility. Across nominal load factors of 40\%, 60\%, and 80\%, gains remain positive at 20.06, 9.39, and 8.47 pp. The 1.5B deployment records positive gains on both L4 (9.39 pp) and A100 (4.99 pp), and the A100/Qwen2.5-7B deployment records an 8.28-pp gain.

\textbf{Prototype overhead.} To attribute runtime differences to PrefixShield's added tracking and victim-selection logic, we compare against vLLM's unmodified LRU path under an identical workload and system configuration. At the default 60\% nominal load profile, LRU and PrefixShield sustain approximately 5.955 and 5.956 requests/s, respectively; this measurement checks incremental policy cost under load and is not a saturation-throughput comparison. At the default-scale 512-block/8-group setting, victim selection takes 0.483\,ms on average and 0.511\,ms at P95, keeping the measured Central Processing Unit (CPU) selector cost below one millisecond. Scaling the selector microbenchmark to 4096 blocks and 32 groups raises the corresponding values to 3.080/3.149\,ms, showing that the straightforward software scan becomes measurable at larger candidate sets and motivating indexed per-group/per-segment queues. Python metadata occupies 218.2\,KiB at 512 blocks/8 groups and 1.57\,MiB at 4096 blocks/32 groups; these figures include Python object overhead and represent prototype footprint.

\begin{figure*}[t]
\centering
\includegraphics[width=0.9\textwidth]{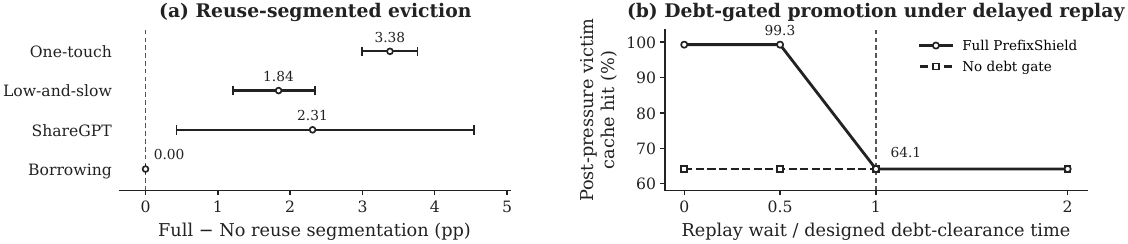}
\caption{Mechanism isolation. (a) Paired cache-hit gain of full PrefixShield over no reuse segmentation, with 95\% bootstrap CIs. (b) Post-pressure victim hit under delayed replay for full PrefixShield and no debt gate; the dashed line marks the designed debt-clearance time.}
\label{fig:mechanism-ablation}
\end{figure*}

\textbf{Reuse-segmented eviction.} Fig.~\panelref{fig:mechanism-ablation}{a} isolates the second stage of victim selection. Full PrefixShield gains 3.38 pp under one-touch (95\% CI: 2.99--3.76), 1.84 pp under low-and-slow (95\% CI: 1.21--2.34), and 2.31 pp on ShareGPT (95\% CI: 0.43--4.54) relative to the no-segmentation variant. Borrowing is unchanged at 0 pp because no eviction differentiation is needed when idle capacity is available. These results separate the two reclamation decisions: responsibility chooses which candidate-owning group supplies victims, whereas segmentation chooses which blocks within that group are exposed first.

\textbf{Debt-gated promotion.} Fig.~\panelref{fig:mechanism-ablation}{b} replaces a purely internal promotion audit with an end-to-end delayed-replay ablation. The no-debt-gate variant retains the same accounting, responsible-group selection, and segmented eviction but allows reuse-eligible blocks to promote regardless of outstanding debt. With zero or half of the designed clearance wait, full PrefixShield promotes none of the 193 replay-eligible blocks and gates all 193, while the ungated variant promotes all 193. Under subsequent neutral pressure, full PrefixShield evicts 74 attacker blocks and no victim blocks, whereas the ungated variant evicts only 2 attacker blocks and 72 victim blocks; the resulting victim cache-hit advantage is 35.16 pp. At the designed clearance-time point, measured replay debt has reached zero, both variants promote all 193 eligible blocks, and their victim hit ratios converge. Debt gating therefore delays promotion to the \textsc{Reuse} class only while admission responsibility remains outstanding: it prevents reuse laundering during debt, but it does not permanently penalize a group after quiescent recovery.

\section{Discussion}
\label{sec:discussion}

\textbf{When persistent responsibility matters.} The comparison with S3-FIFO separates object-level reuse awareness from group-level responsibility. Under direct one-touch churn and two-pass replay, PrefixShield's gains show that object-local reuse history does not identify which accounting group repeatedly creates reclamation pressure. Reuse-aware replacement and responsibility-aware reclamation therefore address complementary questions: the former estimates which objects are worth retaining, while the latter determines which group should supply eviction candidates when shared capacity is contested.

\textbf{Responsibility without rigid partitioning.} Admission debt records accountability for creating pressure rather than an entitlement limit. It neither denies admission nor imposes a fixed occupancy ceiling, and allocation from true-free blocks remains unrestricted; consequently, idle capacity stays borrowable, as Table~\ref{tab:borrowing} demonstrates. Responsibility is also temporary: once excess creation stops and debt clears, the group again becomes eligible for reuse promotion. Because debt accrues independently per accounting group, these properties extend beyond two-group traces: concurrent aggressors each accumulate their own responsibility, and Eq.~\eqref{eq:responsible-group} directs reclamation to the group with the largest projected normalized debt. The prototype associates each materialized block with a stable, operator-defined accounting group while retaining a shared physical block pool, thereby preserving unambiguous responsibility without sacrificing work conservation.

\textbf{Closing the gap between request-time control and persistent state.} Existing serving controls regulate resource use while requests are active, but they do not account for cache pressure that persists after those requests complete. PrefixShield fills this gap by carrying responsibility into the cache-reclamation path, where it determines which accounting group should bear the consequences of continued state creation. This role is distinct from value-aware placement and replacement, which estimate which objects are worth retaining. Deployments requiring strict guarantees can additionally reserve a protected region beneath the shared pool while allowing the remaining capacity to remain work-conserving.

\section{Related Work}
\label{sec:related}

\begin{table}[!h]
\centering
\caption{Representative lines of work by their primary systems question.}
\label{tab:related-positioning}
\footnotesize
{\setlength{\tabcolsep}{2pt}
\begin{tabular}{@{}>{\raggedright\arraybackslash}p{0.18\columnwidth}>{\raggedright\arraybackslash}p{0.29\columnwidth}>{\raggedright\arraybackslash}p{0.44\columnwidth}@{}}
\toprule
Line & Representative systems & Primary question \\
\midrule
KV reuse and placement & Online Context Caching~\cite{gao2025onlinecontext}, Mooncake~\cite{qin2025mooncake}, MELL~\cite{liu2025mell} & Where should reusable KV state reside, move, or be provisioned? \\
\addlinespace[3pt]
Replacement and admission & S3-FIFO~\cite{yang2023s3fifo}, TinyLFU~\cite{einziger2017tinylfu} & Which objects are valuable enough to retain or admit? \\
\addlinespace[3pt]
Service and cache sharing & Virtual Token Counter~\cite{sheng2024fairness}, FairRide~\cite{pu2016fairride}, NyxCache~\cite{wu2022nyxcache} & How should service and cache resources be shared or isolated across clients? \\
\addlinespace[3pt]
\textbf{PrefixShield} & \textbf{This work} & \textbf{Who remains responsible for persistent shared-state pressure after admission?} \\
\bottomrule
\end{tabular}}
\end{table}

KV-management systems optimize how reusable state is represented, placed, moved, or provisioned across conversations, devices, and memory tiers~\cite{kwon2023pagedattention,zheng2024sglang,yu2025pensieve,yao2025cacheblend,yu2025iccache,qin2025mooncake,gao2025onlinecontext,liu2025mell,liu2024cachegen,jiang2025thunderserve,zhang2025jenga,li2025oneiros,yang2025lserve,zhang2026edgekv}. These mechanisms reduce computation or data movement, but they do not preserve which accounting group created later reclamation pressure. S3-FIFO~\cite{yang2023s3fifo} and TinyLFU~\cite{einziger2017tinylfu} use object-value evidence for replacement or admission, while FairRide and NyxCache study how shared cache capacity is divided, isolated, or limited across clients~\cite{pu2016fairride,wu2022nyxcache}. As Table~\ref{tab:related-positioning} summarizes, PrefixShield contributes a complementary lifecycle property: responsibility for new persistent state survives admission, gates reuse promotion while debt remains, and selects which accounting group bears reclamation pressure.

\section{Conclusions and Future Work}
\label{sec:conclusion}
Shared LLM prefix caches create persistent state whose future reclamation cost may fall on groups other than those that admitted it. PrefixShield closes this admission-responsibility gap by metering newly materialized full KV blocks, gating reuse promotion while responsibility remains outstanding, and combining projected responsibility with reuse-segmented victim selection. Our evaluation shows that persistent responsibility addresses a failure mode left unresolved by object-value replacement: direct churn and replay can shift reclamation cost away from the groups that created the cache pressure. PrefixShield improves victim cache-hit rate by 8.64 percentage points over reuse-aware replacement under one-touch pollution and by 2.00 percentage points under two-pass replay, remains statistically indistinguishable under benign ShareGPT traffic, and preserves work-conserving access to idle capacity. Its selector overhead remains below one millisecond at the 95th percentile, with no measurable throughput loss. Object value ranks \emph{what to retain}; persistent responsibility selects \emph{which group bears reclamation pressure}. Future work includes extending responsibility accounting to cross-group content deduplication, which requires multi-principal ownership and admission rules, and replacing the prototype's software scan with indexed per-group and per-segment queues in a systems-language implementation.

\bibliographystyle{IEEEtran}
\bibliography{PrefixShield_references}

\end{document}